\documentclass[9pt, sigconf,screen=true,bookmarks=false]{acmart}
\usepackage{algorithmic}
\usepackage{graphicx}
\usepackage{textcomp}
\usepackage{xcolor}

\copyrightyear{2022} 
\acmYear{2022} 
\setcopyright{rightsretained}
\acmConference[ICCAD '22]{Proceedings of the 2022 International Conference on Computer-Aided Design (ICCAD)}{October 30-- November 3, 2022}{San Diego, CA, USA}
\acmBooktitle{Proceedings of the 2022 International Conference on Computer-Aided Design (ICCAD) (ICCAD '22), October 30-- November 3, 2022, San Diego, CA, USA}

\definecolor{citecolor}{RGB}{34,139,34}
\definecolor{mydarkblue}{rgb}{0,0.08,1}
\definecolor{mydarkgreen}{rgb}{0.02,0.6,0.02}
\definecolor{mydarkred}{rgb}{0.8,0.02,0.02}
\definecolor{mydarkorange}{rgb}{0.40,0.2,0.02}
\definecolor{mypurple}{RGB}{111,0,255}
\definecolor{myred}{rgb}{1.0,0.0,0.0}
\definecolor{mygold}{rgb}{0.75,0.6,0.12}
\definecolor{myblue}{rgb}{0,0.2,0.8}
\definecolor{mydarkgray}{rgb}{0.,0.2,0.2}

\definecolor{lightred}{RGB}{255,235,235}
\definecolor{lightgreen}{RGB}{235,255,235}
\definecolor{lightblue}{RGB}{235,235,255}
\definecolor{lightcyan}{RGB}{235,255,255}
\definecolor{lightmagenta}{RGB}{255,235,255}
\definecolor{lightyellow}{RGB}{255,255,235}

\definecolor{qxkcolor}{RGB}{215,235,255}
\definecolor{softmaxcolor}{RGB}{230,235,255}
\definecolor{probxvcolor}{RGB}{255,255,235}

\definecolor{topkcolor}{RGB}{255,235,235}
\definecolor{zecolor}{RGB}{255,255,235}
\definecolor{dynacolor}{RGB}{235,255,255}

\definecolor{reviewcolor}{RGB}{0,0,200}

\renewcommand\footnotemark{}

\newcommand{\tq}{TorchQuantum\xspace}

\newcommand{\etc}{{etc.}\xspace}

\newcommand{\x}{$\times$\xspace}

\newcommand{\nisq}{NISQ\xspace}

\newcommand{\vs}{vs.\xspace}
\newcommand{\cnot}{\texttt{CNOT}}

\newcommand{\rz}{\texttt{RZ}\xspace}

\newcommand{\sx}{\texttt{SX}}

\newcommand{\xgate}{\texttt{X}}

\newcounter{rlabelno}

\usepackage{color,xcolor}
\usepackage{xspace}
\usepackage{multirow}
\usepackage{booktabs}
\usepackage{lipsum}
\usepackage{soulutf8}
\usepackage{diagbox}
\usepackage{physics}

\usepackage{colortbl}

\usepackage{enumitem}
\usepackage{braket}

\usepackage{amsmath}
\usepackage{amsfonts}
\usepackage{url}

\usepackage{afterpage}
\usepackage{ulem}
\usepackage[ruled,vlined]{algorithm2e}
\usepackage{setspace}

\usepackage{pifont}

\usepackage{titlesec}

\hypersetup{
    colorlinks=true,
    linkcolor=magenta,
    filecolor=magenta,   
    citecolor=citecolor,
    urlcolor=blue,
}

\usepackage{soul}

\begin{document}

\pagestyle{plain}
\title{QuEst: Graph Transformer for \underline{Qu}antum Circuit Reliability \underline{Est}imation}
\subtitle{TorchQuantum Case Study for Robust Quantum Circuits}

\author{Hanrui Wang$^1$, Pengyu Liu$^2$, Jinglei Cheng$^3$, Zhiding Liang$^4$, Jiaqi Gu$^5$, Zirui Li$^6$, Yongshan Ding$^7$, Weiwen Jiang$^8$, Yiyu Shi$^4$, Xuehai Qian$^3$, David Z. Pan$^5$, Frederic T. Chong$^9$, Song Han$^1$\\
  \small{$^1$MIT $^2$Carnegie Mellon University $^3$ Purdue Univ. $^4$Univ. of Notre Dame $^5$Univ. of Taxes at Austin $^6$Rutgers Univ. $^7$Yale Univ.\\$^8$George Mason Univ. $^9$Univ. of Chicago}\\
  \texttt{\url{https://qmlsys.mit.edu}}
}

\begin{abstract}

Quantum Computing has attracted much research attention because of its potential to achieve fundamental speed and efficiency improvements in various domains. Among different quantum algorithms, Parameterized Quantum Circuits (PQC) for Quantum Machine Learning (QML) show promises to realize quantum advantages on the current Noisy Intermediate-Scale Quantum (NISQ) Machines. Therefore, to facilitate the QML and PQC research, a recent python library called \tq has been released. It can construct, simulate, and train PQC for machine learning tasks with high speed and convenient debugging supports.
Besides quantum for ML, we want to raise the community's attention on the reversed direction: ML for quantum. Specifically, the \tq library also supports using data-driven ML models to solve problems in quantum system research, such as predicting the impact of quantum noise on circuit fidelity and improving the quantum circuit compilation efficiency.

This paper presents a case study of the ML for quantum part in \tq. Since estimating the noise impact on circuit reliability is an essential step toward understanding and mitigating noise, we propose to leverage classical ML to predict noise impact on circuit fidelity. Inspired by the natural graph representation of quantum circuits, we propose to leverage a \textit{graph transformer} model to predict the noisy circuit fidelity. We firstly collect a large dataset with a variety of quantum circuits and obtain their fidelity on noisy simulators and real machines. Then we embed each circuit into a graph with gate and noise properties as node features, and adopt a graph transformer to predict the fidelity. We can avoid exponential classical simulation cost and efficiently estimate fidelity with polynomial complexity.

Evaluated on 5 thousand random and algorithm circuits, the graph transformer predictor can provide accurate fidelity estimation with RMSE error \textbf{0.04} and outperform a simple neural network-based model by \textbf{0.02} on average. It can achieve \textbf{0.99} and \textbf{0.95} R$^2$ scores for random and algorithm circuits, respectively. Compared with circuit simulators, the predictor has over \textbf{200\x} speedup for estimating the fidelity. The datasets and predictors can be accessed in the \href{https://github.com/mit-han-lab/torchquantum}{\tq} library.

\end{abstract}

\maketitle
\pagenumbering{gobble}

\section{Introduction}

Quantum Computing (QC) presents a new computational paradigm that has the potential to address classically intractable problems with much higher efficiency and speed. 
It has been shown to have an exponential or polynomial advantage in various domains such as combinatorial optimization~\cite{farhi2014quantum}, molecular dynamics~\cite{peruzzo2014variational, liang2022pan}, and machine learning~\cite{biamonte2017quantum, rebentrost2014quantum, liang2021can,liang2022variational,jiang2021co, cheng2020accqoc,hu2022quantum, wang2021exploration}, \etc
By virtue of breakthroughs in physical implementation technologies, QC hardware has advanced quickly during the last two decades. Multiple QC systems with up to 127 qubits have been released recently~\cite{ibm127, rigetti, google72, intel49}.

\begin{figure}[t]
    \centering
    \includegraphics[width=\columnwidth]{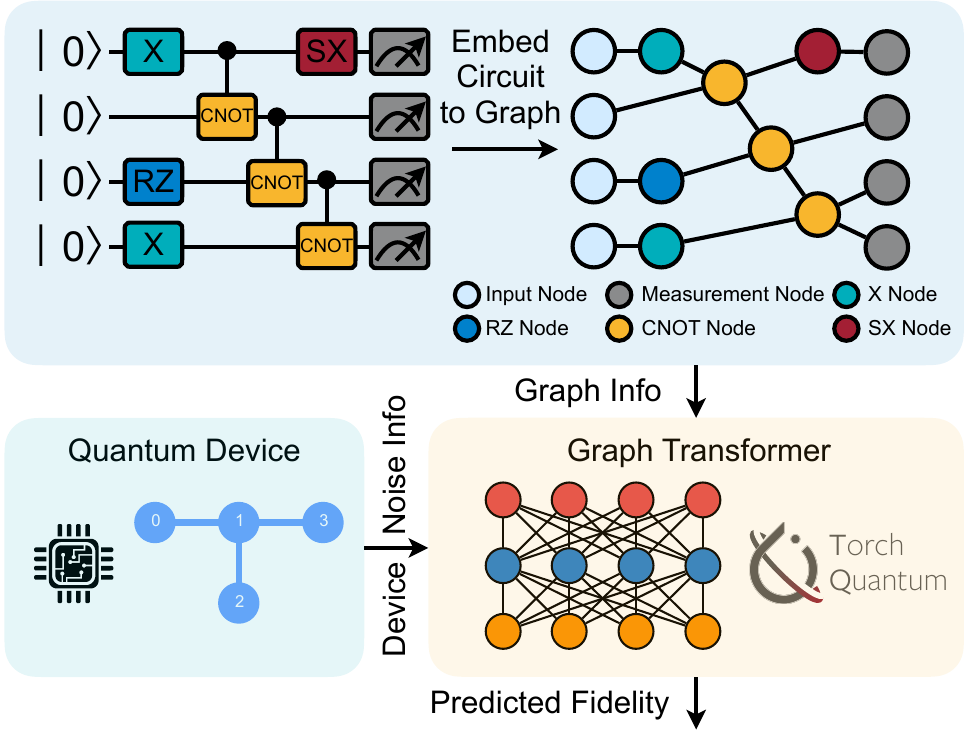}
    \vspace{-20pt}
    \caption{The proposed fidelity prediction framework. The quantum circuit is firstly embedded into a graph in which the nodes are gates and edges are execution orders. The feature vector on each node contains the device noise information, such as gate error rates. The graph is processed by a graph transformer in \tq to estimate circuit fidelity.}
    \label{fig:teaser}
    \vspace{-10pt}
\end{figure}

Despite the promising developments, it is still anticipated that before we enter the fault-tolerant era, we will spend a number of years in the Noisy Intermediate Scale Quantum (\nisq)~\cite{preskill2018quantum} stage. In this stage, the qubits and quantum gates suffer from significant error (around $10^{-3}$), which is the bottleneck towards quantum advantages. Therefore, Parameterized Quantum Circuits (PQC) have attracted increasingly more attention thanks to their flexibility in the circuit architecture (ansatz) and parameters that provides vast space for noise mitigation and optimizations.

To facilitate the robust quantum circuits, especially parameterized quantum circuits for quantum machine learning, the \tq library is released, which supports easy construction, simulation, and fast parameter training of PQCs. Several noise mitigation techniques, such as noise-aware ansatz search~\cite{wang2022quantumnas}, noise-aware parameter training~\cite{wang2021roqnn}, gradient pruning for robust on-chip training~\cite{wang2022chip}, are also supported in the library.

Although plenty of work has been focusing on quantum for machine learning with parameterized circuits, little research explores another direction -- using machine learning to solve quantum system research problems. To fill this vacancy, the \tq library also provides multiple classical machine learning models to perform quantum compilation, reliability estimation tasks, \etc

\begin{figure}[t]
    \centering
    \includegraphics[width=\columnwidth]{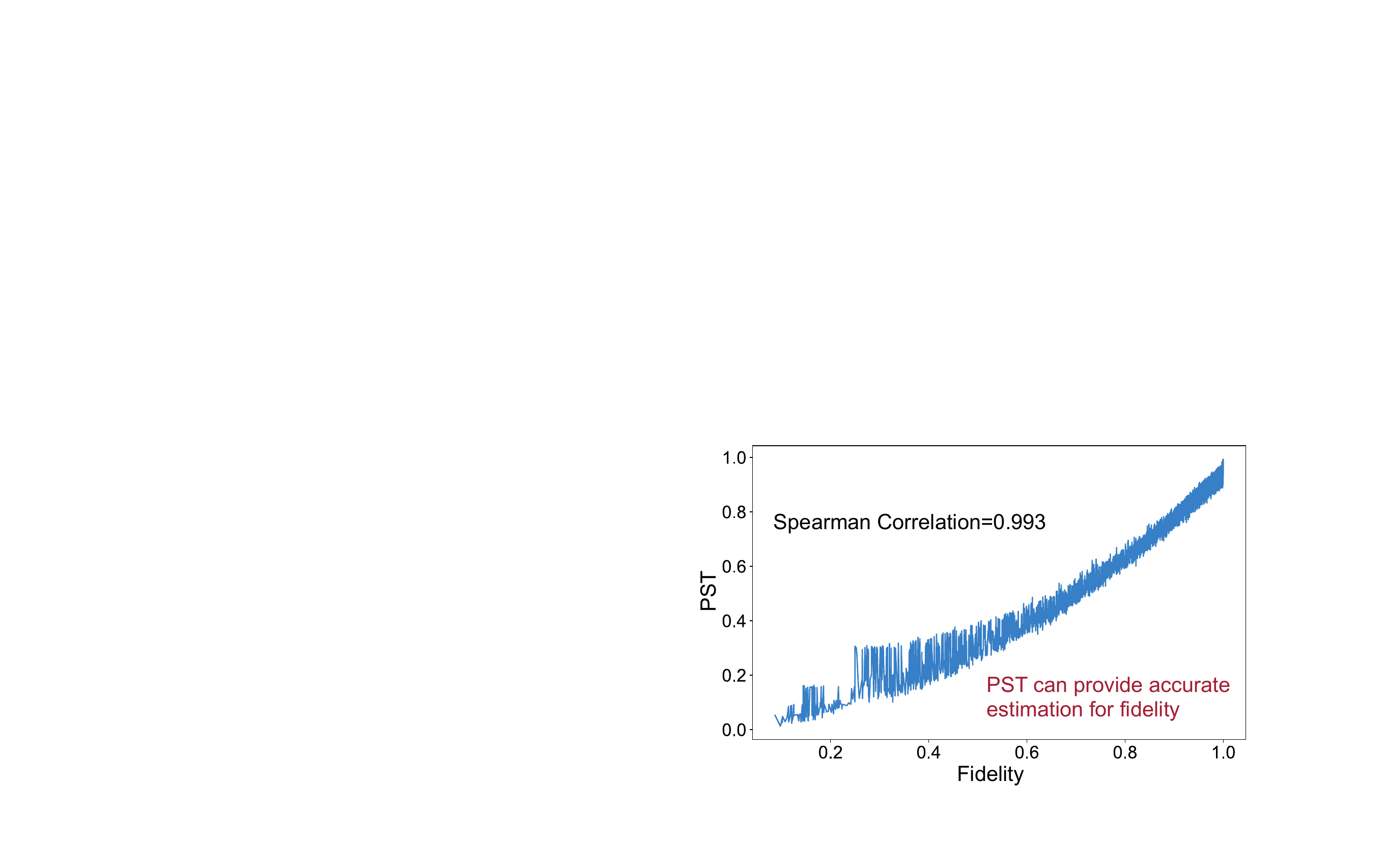}
    \vspace{-20pt}
    \caption{Relationship between fidelity and PST of random circuits. The PST of a circuit is obtained by appending the inverse circuit to the original one and executing. There is a strong positive correlation (Spearman = 0.993) between the two metrics, so it is sufficient for the predictor to output PST.}
    \label{fig:fid_pst}
    \vspace{-10pt}
\end{figure}

In this paper, we show one case study of machine learning for quantum -- using graph transformer models to estimate the quantum circuit fidelity under noise impact, as shown in Figure~\ref{fig:teaser}. Due to the limited quantum resources, it is highly desirable to estimate the circuit performance before submitting it for execution. If the fidelity of a circuit is lower than a threshold, running it on real quantum machines will not generate any meaningful result. One straightforward method is to perform circuit simulation on noisy simulators, but the exponentially increasing cost is prohibitive for circuits with many qubits. Therefore, in this work, we propose a polynomial complexity method in which a data-driven graph transformer is trained to perform fidelity estimation. Intuitively, estimating the fidelity does not require precisely computing the complete density matrix. So there are opportunities that the data-driven method can provide accurate enough estimation with low computation costs. In fact, there have been works on predicting circuit reliability using simple machine learning models~\cite{9251243}. However, it considers neither any graph information of the circuit nor the noise information and thus has less accurate predictions in experimental results.

The first step of the framework is to collect a large dataset containing various randomly generated circuits and circuits from common quantum algorithms. We run the circuits on both noisy simulators and real quantum machines. On simulators, we change the properties of the qubits, such as T1 and T2, and the error rates of gates to diversify the data samples. The dataset contains over 20 thousand samples on simulators and 25 thousand samples on real quantum machines. In order to reduce the overhead of collecting a dataset, we use the ``Probability of Successful Trials" (PST)~\cite{tannu2019not} as the proxy for the fidelity following the setting in~\cite{9251243}. Specifically, for each circuit, we will concatenate the inverse of the circuit to the original one and execute. Since the original quantum state is all zero, the ground truth output of the concatenated circuit will still be all zero. Therefore, the PST will be the frequency of getting all zero bit-string. 
The dataset is embedded in the \tq library and can be easily accessed for future studies.

Secondly, motivated by the fact that \textit{quantum circuits are graphs}, we propose to leverage a graph transformer to process the circuit information. The nodes of the graph are the quantum gates, input qubits, and measurements. The edges are determined by the sequence of gate executions. The feature vector on each node contains gate type, qubit index, qubit T1, T2 time, gate error rate, \etc, to capture operation and noise information. In one layer of the graph transformer, the attention layer will capture the correlations between each node and its neighbors according to the graph and compute the updated feature vector. Several fully-connected layers are appended at the end to regress the circuit PST.

Overall, we present a case study on using graph transformer models in the \tq library to estimate circuit fidelity under noise. The contributions are summarized as below:
\begin{itemize}
    \item \textbf{A dataset for circuit fidelity} on various noisy simulators and real machines is presented and embedded in the \tq library to facilitate research on reliability estimations. It contains 20K simulation samples and 25K real machine samples.
    \item \textbf{A graph transformer model} is constructed and trained to process the quantum circuit graph and feature vectors on nodes to provide accurate fidelity prediction.
    \item \textbf{Extensive evaluations} on around 2 thousand circuits on noisy simulators and 3 thousand circuits on real machines demonstrate the high accuracy of the predictor. It achieves \textbf{0.04} RMSE and over \textbf{0.95} R$^2$ scores with \textbf{200\x} speedup over circuit simulators. 
\end{itemize}

\section{Related Work}

\subsection{Quantum Basics}
A quantum bit (qubit) can be in a linear combination of the two basis states 0 and 1, in contrast to a classical bit, $\ket{\psi} = \alpha\ket{0} + \beta\ket{1},$ for $\alpha,\beta\in\mathbb{C},$ where $|\alpha|^2+|\beta|^2=1$. Only one of the $2^n$ states can be stored in a classical $n$-bit register. 
However, we can employ an $n$-qubit system to describe a linear combination of $2^n$ basis states due to the ability to build a superposition of basis states. To perform computation on a quantum system, we use a \textit{quantum circuit} to manipulate the state of qubits. A given quantum system can be expressed as a Hamiltonian function and solved by Schr{\"o}dinger's equation, and these operational steps can be performed by various \textit{quantum gates}.
Results of a quantum circuit are obtained by qubit readout operations called \textit{measurements}, which collapse a qubit state $\ket{\psi}$ to either $\ket{0}$ or $\ket{1}$ probabilistically according to the amplitudes $\alpha$ and $\beta$.

\begin{figure*}[t]
    \centering
    \includegraphics[width=\textwidth]{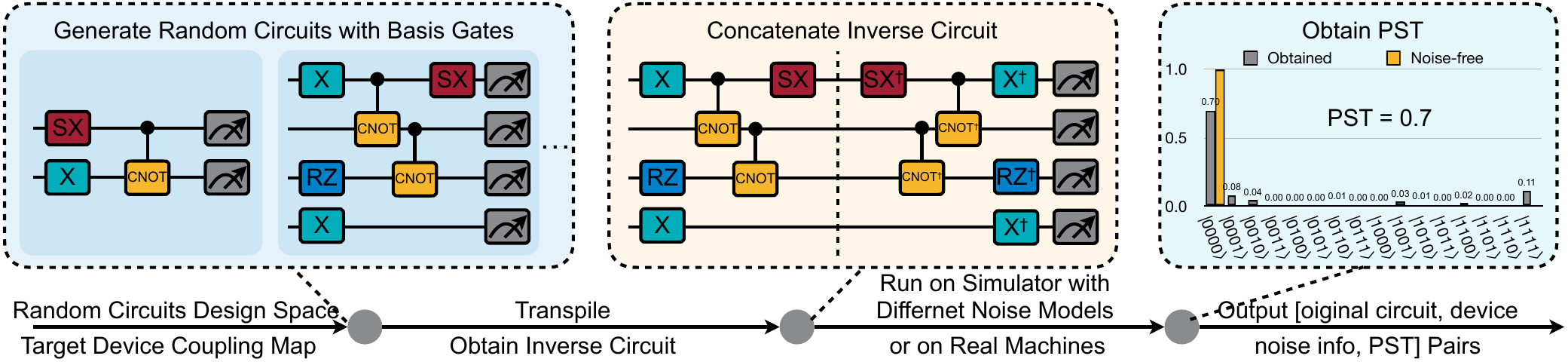}
    \caption{Overview of the dataset generation process. i) Prepare random circuits by mixing basis gate: \rz, \sx, \xgate, \cnot, namely constructing native circuits. ii) Inverse all the gates in the transpiled native circuit and then concatenate the inverse circuit to the original transpiled native circuit. iii) Calculate PST by dividing the number of trials (shots) with all zero state by the number of total trials (shots). 
    }
    \vspace{-10pt}
    \label{fig:dataset_generation}
    
\end{figure*}

\subsection{Quantum Errors}
Quantum errors are one of the most significant challenges that NISQ-era quantum computing experiences. On real quantum machines, 
errors occur because of the interactions between qubits and the environment, control errors, and interference from the environment \cite{krantz2019quantum, bruzewicz2019trapped, magesan2012characterizing}. Qubits undergo \textit{decoherence error} over time, and quantum gates introduce \textit{operation errors} (such as coherent/stochastic errors) into the system. 
These systems need to be characterized~\cite{magesan2012characterizing} and calibrated~\cite{ibm_2021} frequently to mitigate the quantum noise impacts.

The errors seriously interfere with the function of quantum circuits and form obstacles to further optimization of quantum circuits. A number of noise mitigation techniques have been developed to attenuate negative effects~\cite{das2022afs, wang2022quantumnas, wang2021roqnn, liang2021can, hua2021autobraid, ravi2022vaqem, krinner2022realizing, cheng2022topgen}.~\cite{wang2021roqnn} proposes a framework to improve the quantum circuits' robustness by making them aware of noise. It consists of three main techniques: injection of gate errors, regularization, and normalization of measurement outcomes. Another literature~\cite{liang2021can} integrates the gate error characteristics into the mapped quantum circuit to improve robustness.

\subsection{Fidelity Estimation and Prediction}
In order to validate and characterize the states generated by a quantum computer, it is crucial to estimate the fidelity of quantum states~\cite{wang2021quantum,gilyen2022improved}. 
However, calculating fidelities is already quite computationally expensive.
Numerous efforts have been made to address this problem in the past few years.
Variational quantum algorithms have been adopted by recent works to perform fidelity estimation~\cite{chen2021variational,tan2021variational,cerezo2020variational}.
Machine learning-based and statistical methods are also proposed to estimate the fidelity~\cite{zhang2021direct,yu2022statistical,9251243}. 
In addition, ``classical shadow'' is proposed for more efficient tomography~\cite{huang2020predicting}, which can also benefit fidelity estimation.
The works mentioned above present various methods for estimating fidelity.
Fewer works, however, have focused on predicting fidelity given a quantum circuit and a noisy backend.
~\cite{9251243} derives a fidelity prediction model using polynomial fitting and a shallow neural network. 
The noisy backend is considered as a black box in that work.
~\cite{nishio2020extracting,tannu2019ensemble} calculate fidelity with a simple equation and use it as a metric to optimize the compilation workflow.
These methods are inaccurate and do not account for the structure of quantum circuits or noisy backends.

\subsection{Randomized Benchmarking}
Plenty of techniques have been developed to estimate the fidelity of quantum circuits and identify errors in \nisq computers, and they can provide indicators of the quality of quantum circuits and directions for further improvement of quantum hardware. Among them, randomized benchmarking is the most prominent~\cite{knill2008randomized,magesan2011scalable,magesan2012characterizing} one. Randomized benchmarking can estimate the fidelity of certain gates or circuits and further characterize noises to very high accuracy in the presence of state preparation and measurement errors. However, randomized benchmarking has several limitations. For example, it usually requires strong assumptions about the error pattern, such as assuming the errors are gate-independent, and the benchmarked gate set must have group structures.

\subsection{Transformers}
The attention~\cite{bahdanau2014neural, NIPS2017_3f5ee243} based Transformer models~\cite{wang2021spatten, wang2020efficient} have prevailed in sequence modeling. Recently, it is also widely applied in other domains such as vision transformer~\cite{dosovitskiy2020image} for computer vision and graph transformer (graph attention networks) ~\cite{velivckovic2017graph, yun2019graph, 9218757, wang2018learning} for graph learning. The graph transformer leverages the attention mechanism to generate the updated features of the next layer for each node. The Query vectors come from the center node, while the Key and Value vectors are calculated from the neighboring nodes. Recently, several variants of traditional transformers have been proposed, including AGNN, which removes all the FC linear layers in the model~\cite{thekumparampil2018attention}, Modified-GAT~\cite{ryu2018deeply}, which proposes gate-augmented attention for better feature extraction, Linear Attention~\cite{shen2021efficient}, which reduces the complexity of attention to linear cost, and Hardware-Aware Transformer~\cite{wang2020hat} that adjusts the architecture according to the hardware latency feedback.

\section{Circuit Fidelity Dataset}
In classical computing, training datasets must be fed into the machine learning algorithms before validation datasets (or testing datasets) can be employed to validate the model's interpretation of the input data.
However, when dealing with the fidelity prediction problem, we do not have an off-the-shelf dataset that can be used to train and evaluate different methods.
To address this problem, we present a scheme for generating datasets and incorporating the gathered datasets into \tq in order to provide relevant researchers with appropriate starting points.

\subsection{Metrics}

\label{metrics}
In order to accurately estimate the ``success rate'' of quantum circuits on noisy devices, the conception of fidelity is introduced, which is a measure of the ``closeness'' of two quantum states.
In noisy quantum computing, fidelity is adopted to illustrate the difference between the quantum states generated by noisy devices and those generated by noiseless classical simulations.
Obtaining the fidelity of quantum circuits is, however, computationally costly -- exponential to the qubit number.
Intricate tomography would be required to ``restore'' or ``describe'' quantum states\cite{huang2020predicting}.
To solve such a problem, we adopt the idea of ``Probability of Successful Trials" (PST)~\cite{tannu2019not} as the proxy of fidelity.
$$PST = \frac{\# Trials\ with\ output\ same\ as\ initial\ state}{\#Total\ trials}$$
Instead of measuring the fidelity of quantum circuits, we count the proportion of unchanged qubits (all zeros) after concatenating the circuits with their inverse.
For concatenated circuits, the proportion will be one if we conduct simulations on a noise-free simulator.
We compare the PST with fidelity for 1400 quantum circuits on simulators.
As shown in Figure~\ref{fig:fid_pst}, they exhibit a strong correlation with a Spearman correlation coefficient of 0.993.
Therefore, we can conclude that PST can provide accurate fidelity estimations.

\begin{figure}[t]
    \centering
    \includegraphics[width=\columnwidth]{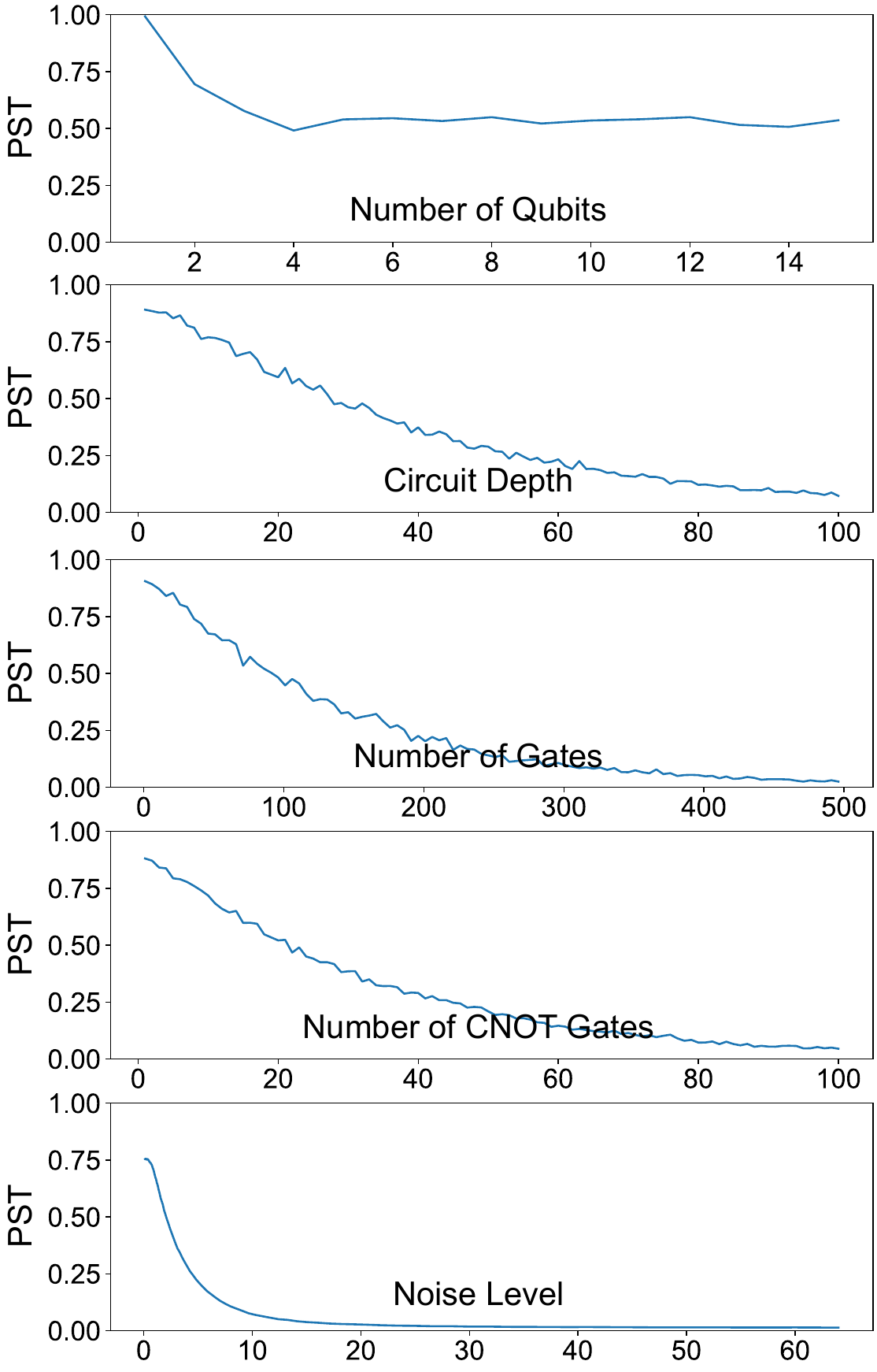}
    \vspace{-20pt}
    \caption{Dataset property profiling.}
    \label{fig:profiling}
    \vspace{-10pt}
\end{figure}
\subsection{Dataset Generation}
\label{data_gen}
As shown in Figure~\ref{fig:dataset_generation}, the generation of random datasets can be broken down into three major steps: initial random circuit generation, concatenation with inverse circuits, and PST calculation.

\textbf{Native Circuit Construction.} 
In the ﬁrst step, random gates are generated from the basis gate set \{\rz, \sx, \xgate, \cnot\} and assigned to quantum circuits to create an initial version of random circuits. 
Single-qubit gates are assigned to all possible qubits, and two-qubit gates are assigned to all available connections in the quantum device. 
After finishing the assignments, the circuits will be compiled to eliminate duplicated gates.
As a result, we consider the number of qubits and gates, the coupling map of quantum devices, and the number of random circuits as parameters during the random circuits generation process.

\textbf{Concatenation of Inverse Circuit.} 
Furthermore, the obtained random circuits will be concatenated with their inverse. The inverse circuit is obtained by reversing the gate sequence of the original circuit and replacing each gate with its inverse gate, as shown in Figure~\ref{fig:dataset_generation} middle.
The purpose of concatenation is to use PST rather than fidelity as our metrics, thereby allowing us to avoid the computationally expensive state tomography.
The detailed reasons are elaborated on in the Section~\ref{metrics}.
For example, assuming a circuit consisting of a \cnot\xspace gate and an \xgate\xspace gate, 
after concatenation, the circuit will be ``\cnot\xspace + \xgate\xspace + barrier + \xgate\xspace + \cnot\xspace''.
A barrier is placed to prevent gate cancellation.
The concatenated circuits will be sent to the backends to obtain PSTs. 
Note that the dataset only contains the original circuits \textit{without} concatenation.
As a result, if we need to evaluate a new quantum circuit, we will feed it to the ML predictor. Then the predicted PST for the concatenated circuit will be returned by the ML model, which is highly correlated with the circuit's fidelity.

\textbf{PST Calculation.} 
The concatenated circuits will then be passed to noisy backends to calculate the PST.
We begin with the default initial state $|00....0\rangle$, and the PST represents the proportion of $|00....0\rangle$ in the output distribution.
Our prediction model takes into account the information from both quantum circuits and noisy backends.
As a result, the quantum circuits are simulated on backends with differing noise levels to create our datasets.
The backends' noise configurations are derived from real NISQ machines, with random constants to change the noise levels.

\begin{figure*}[t]
    \centering
    \includegraphics[width=\textwidth]{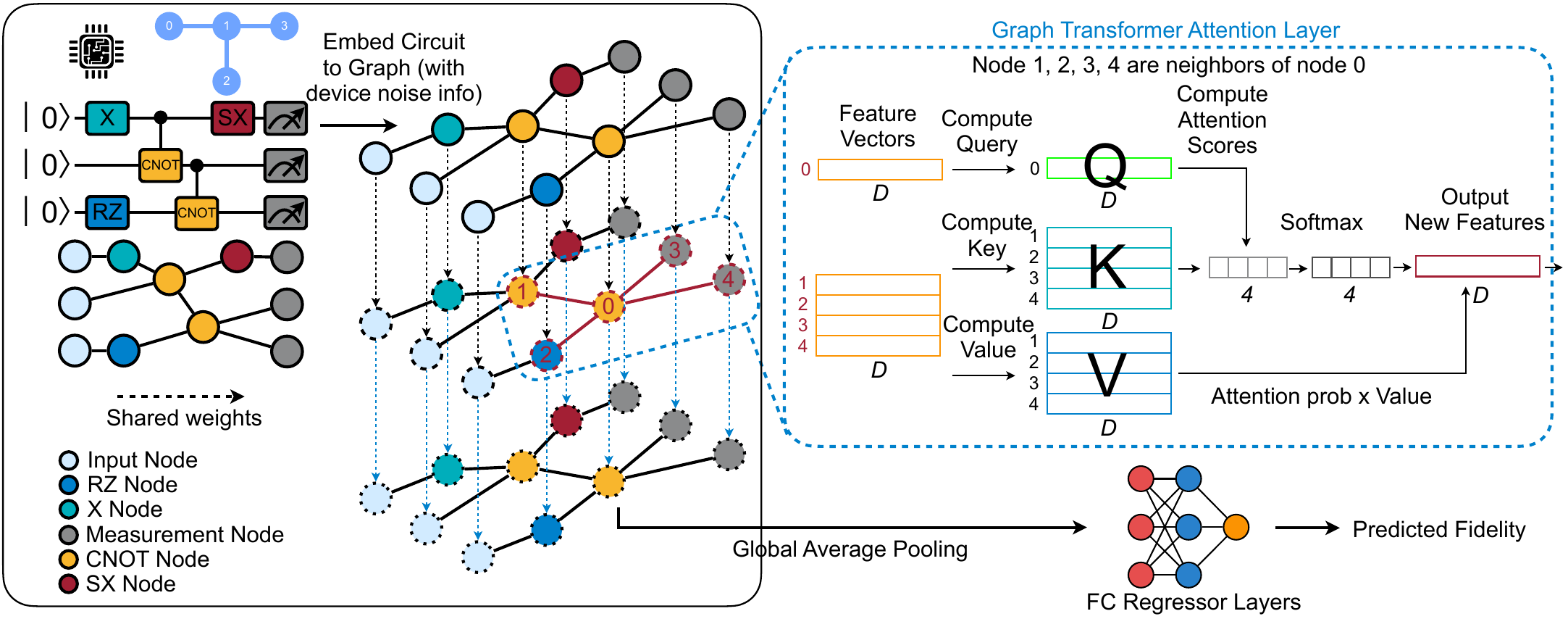}
    \caption{Overview of the Graph Transformer for fidelity prediction. (i) Generate the graph according to quantum circuit, and then generate the feature vector for each of the node according to the quantum device noise information. (ii) For one Graph Transformer layer, we perform graph attention layer to extract information and captures the neighboring correlations. The weight matrices are shared across all nodes. (iii) Finally, a regressor containing several FC layers regresses the circuit PST (an approximation of fidelity).}
    \vspace{-15pt}
    \label{fig:gt}
\end{figure*}

\subsection{Dataset Properties}
\begin{figure*}[t]
    \centering
    \includegraphics[width=\textwidth]{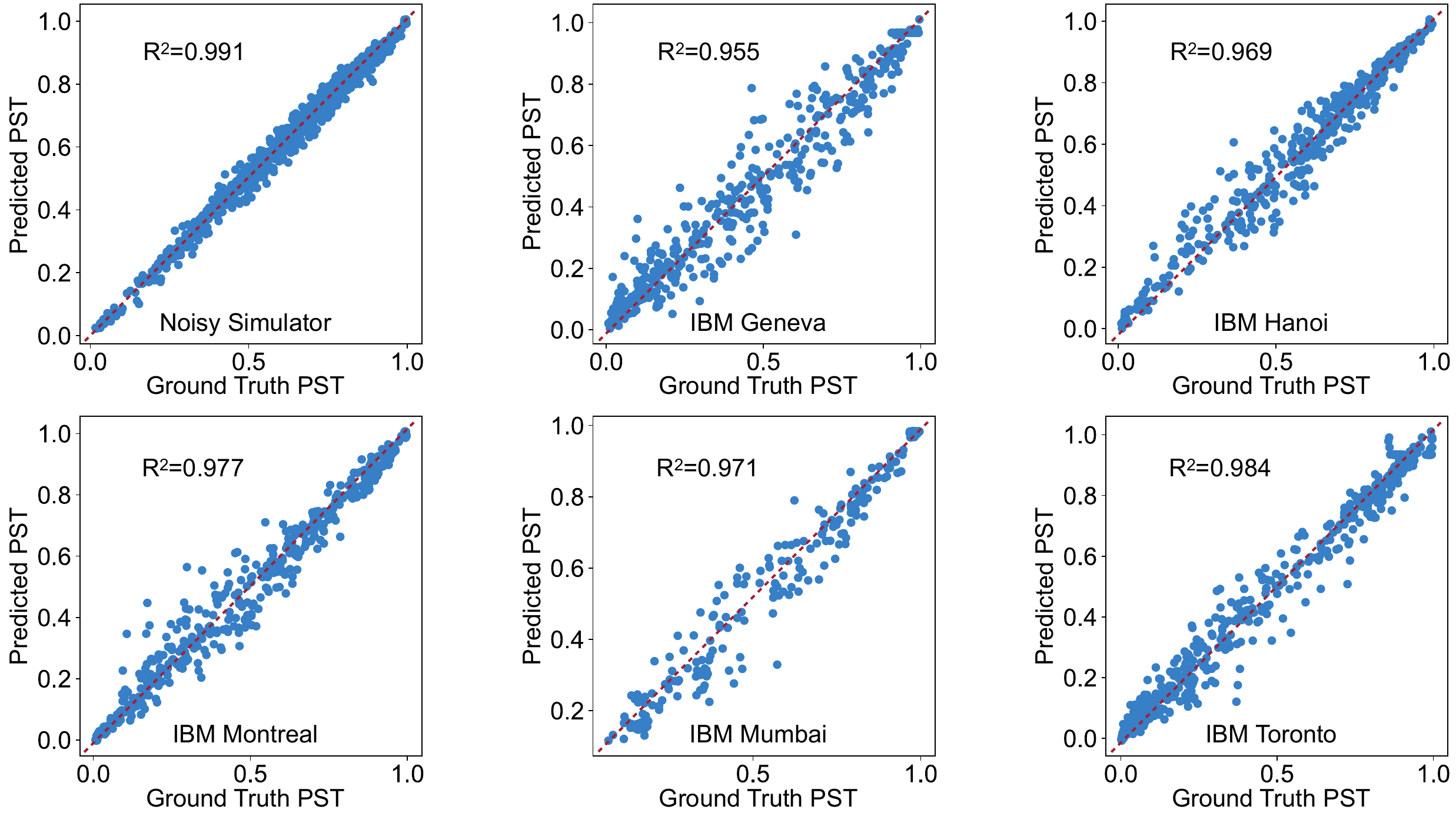}
    \caption{Scatter plots of PST of randomly generated circuits on noisy simulators and 5 real machines. Our transformer can provide accurate estimations of PST with R$^2$ higher than 0.95.}
    \label{fig:scatters}
\end{figure*}

Figure~\ref{fig:profiling} depicts the relationships between PST and various circuit and backend properties.
We can anticipate a lower PST as the number of gates increases.
The PST numbers are also influenced by the number of \cnot\xspace gates, circuit depth, and noise level.
To cover these dimensions, we create random datasets with varying numbers of qubits, gates, and backends of different noise levels.
The PSTs of these circuits are simulated on backends with five different noise levels.
As a result, the random circuits datasets contain 10000 data points on noisy simulators.
We also measure the PSTs of these circuits from five different real NISQ machines. The dataset on real machines contains around 25000 data points.
The performance of our graph transformer model on random circuits is demonstrated in Figure~\ref{fig:scatters}.
In addition, our datasets include circuits used in quantum algorithms such as quantum error correction~\cite{lidar2013quantum}, variational quantum eigensolver~\cite{kandala2017hardware}, Grover search~\cite{grover1996fast}, quantum fourier transform~\cite{coppersmith2002approximate}, quantum approximate optimization algorithm~\cite{farhi2014quantum} and quantum teleportation~\cite{bennett1993teleporting}.
We select a total of 30 circuits derived from quantum algorithms.
The simulations are also carried out on noisy simulators with varying noise levels to collect data points.
The performance of our graph transformer model on these circuits is demonstrated in Figure~\ref{fig:tranditional_scatters}.

\newlength{\oldtextfloatsep}\setlength{\oldtextfloatsep}{\textfloatsep}

\setlength{\textfloatsep}{0pt}%
\begin{algorithm}[!t]
\setstretch{0.75}
\SetKwInOut{Input}{Input}
    \SetAlgoLined
    \textbf{Input:} 
    Circuit graph: $G$ with $K$ nodes\\
    Length of feature vector: $D$\\
    Node features: $\mathbf{H} \in \mathbb{R}^{K \times D}$\\
    Query, Key, Value weights $\{ \mathbf{W}_Q , \mathbf{W}_K, \mathbf{W}_V\} \in \mathbb{R}^{D \times D}$\\
    $\mathbf{Q} = \mathbf{W}_Q \cdot \mathbf{H} $\\
    $\mathbf{K} = \mathbf{W}_K \cdot \mathbf{H} $ \\
    $\mathbf{V} = \mathbf{W}_V \cdot \mathbf{H} $\\
    \SetKwBlock{DoParallel}{do in parallel}{end}
    \DoParallel{
        \For{$i = 0 \mathbf{ \ to \ } K$}{
            Obtain neighbor nodes $\mathcal{N}_i$ according to $G$\\
            $attention\_score_{ij} = \mathbf{Q}_i \cdot \mathbf{K}_j^T, j \in \mathcal{N}_i$\\
            
            $\mathbf{attention\_score} = \mathbf{attention\_score} / $sqrt($|\mathcal{N}_i|$)    \\ 
            $\mathbf{attention\_prob} = $
            Softmax($\mathbf{attention\_score}$)\\
            $\mathbf{attention\_out}_i = \sum_{j\in \mathcal{N}_i} attention\_prob_{ij} \cdot V_j$\\
            $\mathbf{attention\_out}_i \in \mathbb{R}^{D} $
        }
    }
    \textbf{Output:} $\mathbf{attention\_out} \in \mathbb{R}^{K \times D}$\\
    
    \caption{Attention in Graph Transformer}
    \label{algo:attention}

\afterpage{\global\setlength{\textfloatsep}{\oldtextfloatsep}}

\end{algorithm}

\section{Predictor}
The dataset introduced in the previous section enables a data-driven approach to learning the PST from circuit and noise features. This section will continue to present a case study of a deep learning model, graph transformer, for circuit PST prediction. Figure~\ref{fig:gt} shows the overview of the framework. A gate graph is firstly extracted from the circuit. Then the node features are generated according to the gate type, noise information, etc. Next, a graph transformer containing attention operations is introduced to process the node features and neighboring relations. Finally, a PST regression layer outputs the predicted values.

\begin{figure*}[t]
    \centering
    \includegraphics[width=\textwidth]{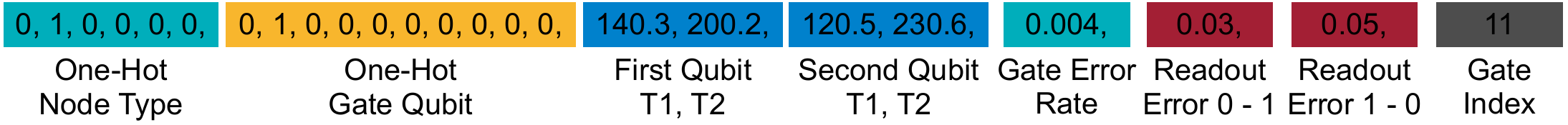}
    \caption{Node feature vector.}
    \label{fig:feature_vector}
\end{figure*}

\subsection{Graph Construction}
We firstly use directed acyclic graphs (DAG) to represent the topology of quantum circuits. Each node represents one qubit, quantum gate, or measurement. Edges represent the time-dependent order of different gates. One example of extracting the graph from the circuit is presented on the left of Figure~\ref{fig:gt}. The connectivity can be encoded into an adjacent matrix. With the \tq framework, the DAG can be conveniently converted from the circuit.

\subsection{Node Features}
For each node in the graph, we generate a vector representing the features. The features include gate type, target qubit index, T1 and T2 of the target qubit, gate error, and gate index, as shown in Figure~\ref{fig:feature_vector}. In our experiments, we set the maximum qubit number to $10$, and the feature vector has a length of $24$. The first $6$ numbers are one-hot vectors describing the gate type: initial input qubit, measurement, \rz, \xgate, \sx, or \cnot. Then we use $10$ numbers to describe the target gate qubit(s). If this gate acts on the $i^\mathrm{th}$ qubit, the $i^\mathrm{th}$ number of the vector is set to $1$ and otherwise $0$. That also applies to multi-qubit gates. Then we use the following $7$ numbers to describe the calibration information of the backend with the following format: [T1, T2 for the first target qubit, T1, T2 for the second target qubit, gate error rate, readout error10, readout error01]. If a feature is not applicable for a particular node, the corresponding value is set to $0$. For example, \rz acts on only one qubit, so T1 and T2 for the second target qubit are set to $0$. Since \rz is not a measurement, readout error10 and readout error01 are set to $0$ also. The last number is used to encode the index of the node. The whole featur vector is illustrated in Figure~\ref{fig:feature_vector}.

\subsection{Graph Transformer}

To process graphs with node features, we propose to use a graph transformer as shown in Figure~\ref{fig:gt} right. The transformer contains multiple layers, each containing the attention operation. The attention is described in Algorithm~\ref{algo:attention}. the Query, Key, and Value vectors for each node are computed with shared weights. Then for one node, we fetch the Key vectors of its neighboring nodes and compute Query $\times$ Key$^T$. The outputs are attention scores which are then normalized according to the square root of the number of neighbors. Softmax is adopted to normalize the attention scores. The output is called attention probability because the values add up to one. The probability vector is then employed as weights to perform a weighted sum of the Value vectors of the neighboring nodes. The output has the same dimension as the input feature of the center node. After that, we perform a residual connection between input and output of attention with a layer normalization. The output will be the feature vector of the next layer. Note that computations on all nodes are done \textit{simultaneously.}

After multiple transformer layers, we obtain a learned feature on each node, with its neighbors influenced. If deep enough, each node can access to features of all nodes in the graph. Finally, we perform a global average pooling of the node features and obtain an aggregated node feature vector. Then a regressor with three FC layers is appended to output the final regressed PST. Besides node feature, we also leverage \textit{global features}, representing the circuit depth, width, and counts of \rz, \xgate, \sx, and \cnot\xspace gates. The global feature vector is concatenated with the aggregated node feature vector and fed to the regressor.

The computational complexity of the proposed graph transformer is polynomial to qubit number since the overall number of gates is typically polynomial to qubit number.

\section{Evaluation}

\subsection{Evaluation Methodology}
\begin{figure*}[t]
    \centering
    \includegraphics[width=\textwidth]{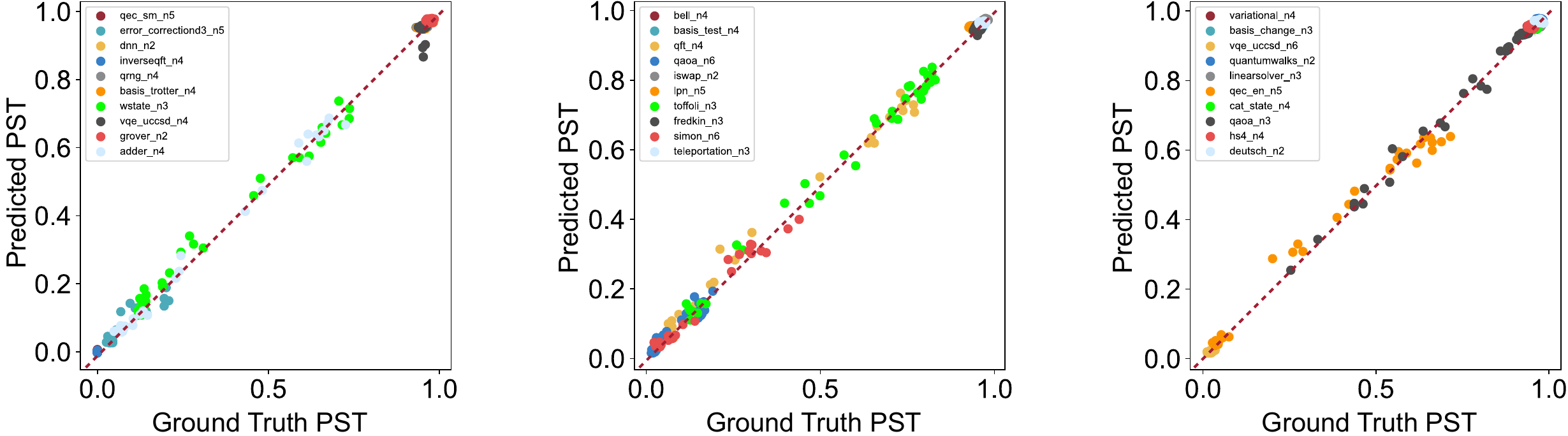}
    \caption{Scatter plots of circuit PST of 30 quantum algorithms on noisy simulators. Our transformer can provide accurate estimations of PST with R$^2$ 0.99.}
    \label{fig:tranditional_scatters}
\end{figure*}

\begin{figure}[t]
    \centering
    \includegraphics[width=\columnwidth]{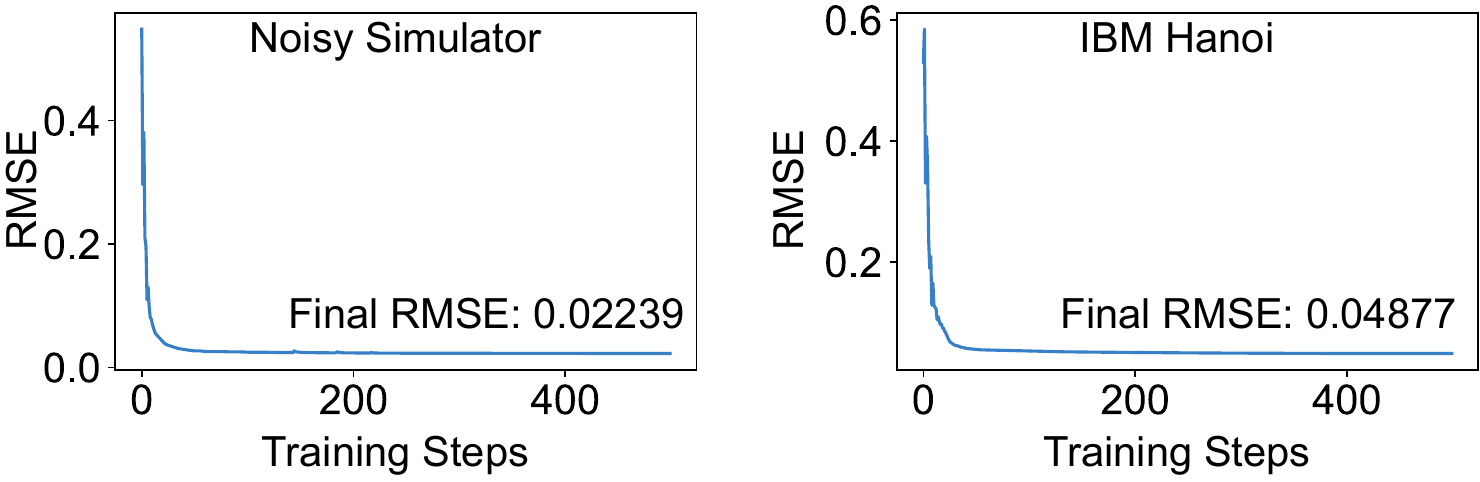}
    \caption{Training curves of transformer models on noisy simulators and IBM Hanoi datasets for random circuits.}
    \label{fig:curves}
\end{figure}

\textbf{Model and Training Setups.}
In the default setup, we use two layers of graph transformers. The embedding dimension is $24$ since we have 24 features. The dimension for the Query, Key, and Value vectors is also $24$. We use single-head attention layers. The global average pooling across nodes generates a single $24$ dimensional vector as the aggregated feature for a circuit. If global features are enabled, we use two FC layers with hidden and output dimensions of $12$ to pre-process and concatenate it with the aggregated node feature. The concatenated feature is processed with additional three FC layers with hidden dimension $128$ and output dimension $1$. This output is treated as the predicted PST value. We use ReLU activation.

We normalize the node features across the dataset by removing the mean and dividing the standard deviation. We then train the models with Adam optimizer for $500$ epochs with a constant learning rate of $10^{-2}$, weight decay $10^{-4}$, batch size $2500$ and MSE loss. Then we choose the model that performs best on the validation set to test on the test set.

\begin{table}[t]
\centering
\renewcommand*{\arraystretch}{1}
\setlength{\tabcolsep}{3pt}
\caption{Prediction RMSE vs. Whether Using Global Features}

\begin{tabular}{l|ccc}
\toprule
Features & Noisy Simulator & IBM Geneva & IBM Hanoi\\
\midrule
\midrule
w/o Global Features & 0.0239 & 0.0757 & 0.0506 \\
\midrule
w/ Global Features & \textbf{0.0232} & \textbf{0.0723} & \textbf{0.0500} \\
\bottomrule
\end{tabular}
\label{tab:rmse_vs_global_feature}

\end{table}

\begin{figure}[t]
    \centering
    \includegraphics[width=\columnwidth]{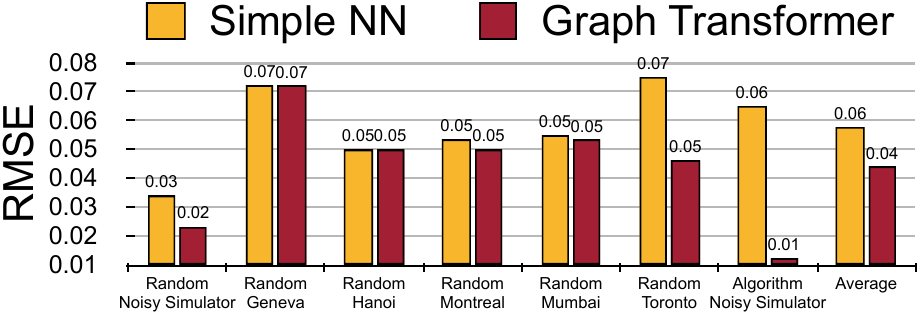}
    \caption{The proposed graph transformer-based model can outperform the simple NN model on various benchmarks.}
    \label{fig:histo}
\end{figure}

\textbf{Dataset Setup.}
For noisy simulators datasets, we have 10000 random circuits and 350 circuits for 30 quantum algorithms each. For real machine datasets, we collect 5000, 5000, 5450, 2750, and 6750 random circuits for IBM Geneva, IBM Hanoi, IBM Montreal, IBM Mumbai, and IBM Toronto, respectively. We split the dataset into three parts, the training set includes $70\%$ data, the validation set includes $20\%$ data, and the test set consists of the last $10\%$.

\subsection{Experimental Results}

Figure~\ref{fig:scatters} shows the scatter plots of transformer predicted PST \vs the ground truth PST for randomly generated circuits on the test set. The red dash line is the $y=x$ line. We train one separate model for each of the backend settings. For results on noisy simulators, the points are close to the $y=x$ line with an R$^2$ value of 0.991. On real machines, the difficulty is greater than on noisy simulators. Although the predictor R$^2$ is lower than noisy simulators, they are still higher than 0.95. Furthermore, as in Figure~\ref{fig:tranditional_scatters}, we select 30 representative quantum algorithms as benchmarks and show the scatter plots for predicted PST on the test set. Each color represents one algorithm circuit under different noise models. We train one common model for the 30 algorithm circuits. The transformer model can effectively track the PST value, especially for those spanning a wide range of PST. The overall R$^2$ for 30 benchmarks is 0.9985. We also show the two representative training curves on noisy simulators and the real quantum machine IBM Hanoi in Figure~\ref{fig:curves}. The training loss converges after around 200 steps. The convergence speed on real machine data is slightly slower than the noisy simulator data and has a higher final RMSE (around 0.05).

Besides, we also compare our transformer-based model with the simple NN model adapted from~\cite{9251243} as in Figure~\ref{fig:histo}. The simple NN model only takes 116 features as input, which include circuit depth, width, and counts of \rz, \xgate, \sx, and \cnot\xspace gates, single-qubit gate counts on each qubit, and two-qubit gate counts on each qubit pair. It uses $3$ FC layers with hidden dimension 128 and ReLU activation to regress the PST. We compare the RMSE on the test set for random circuits on 6 benchmarks and 30 algorithm circuits on noisy simulators. On average, the RMSE of the transformer model is 0.02 better than the simple NN model. On the algorithm circuit, the gap is even more apparent -- up to 0.05. The R$^2$ on algorithm circuits with transformer is also much higher than simple NN (0.9985 \vs 0.9110). That shows the effectiveness of involving circuit graph information in the model.

\begin{table}[t]
\centering
\renewcommand*{\arraystretch}{1}
\setlength{\tabcolsep}{3pt}
\caption{Importance Comparison of Node Features}

\begin{tabular}{l|ccc}
\toprule
Features & Noisy Simulator & IBM Geneva & IBM Hanoi\\
\midrule
\midrule
All Features & 0.0232 & 0.0723 & 0.0500 \\
\midrule
\xspace w/o Gate Error Rate & 0.0235 & 0.0732 &0.0501 \\
 \midrule
\xspace w/o Gate Index & \textbf{0.0247} & 0.0730 & 0.0497  \\
 \midrule
\xspace w/o Gate Type & 0.0236 & \textbf{0.0742} & \textbf{0.0512} \\
 \midrule
\xspace w/o Qubit Index & \textbf{0.0239} & \textbf{0.0736} & \textbf{0.0514} \\
 \midrule
\xspace w/o T1\&T2 & 0.0239 & 0.0707 & 0.0491 \\
\bottomrule
\end{tabular}
\label{tab:rmse_vs_feature}

\end{table}

\subsection{Analysis}
In Table~\ref{tab:rmse_vs_global_feature}, we show the effectiveness of concatenating the global features to the aggregated node features. Adding global features can reduce the RMSE loss on the test set with negligible computational overhead. The effectiveness is especially significant in IBM Geneva, where the RMSE is reduced by around 0.003. 

Table~\ref{tab:rmse_vs_feature} further performs an ablation study on the importance of each feature in the node feature vectors. We remove one feature while keeping all other features in each experiment and then train the model again to obtain the results and report the RMSE loss on the test set. The bold values mark the largest two losses when removing different features. We can see that removing `Qubit Index' severely degrades the accuracy in all three backends. This may be because the qubit index helps the transformer model know the location of the gate. Removing `Gate Type' also has a substantial negative impact since the model will not know the node type. We also observe that removing some features even improves the accuracy. This only happens on the real machine backend and maybe because of the large fluctuations of noise on the real backend.

\begin{table}[t]
\centering
\renewcommand*{\arraystretch}{1}
\setlength{\tabcolsep}{8pt}
\caption{Prediction RMSE vs. Transformer Layer Number}

\begin{tabular}{l|ccc}
\toprule
\# Layers & Noisy Simulator & IBM Geneva & IBM Hanoi\\
\midrule
\midrule
 1 & \textbf{0.0230} & 0.0720 &  \textbf{0.0491} \\
 \midrule
 2 &  0.0232 & 0.0723 & 0.0500 \\
 \midrule
 3 & 0.0232 & \textbf{0.0719} & 0.0500 \\
\bottomrule

\end{tabular}
\label{tab:rmse_vs_layer}

\end{table}

\begin{table}[t]
\centering
\renewcommand*{\arraystretch}{1}
\setlength{\tabcolsep}{12pt}
\caption{Prediction RMSE vs. Number of Shots}

\begin{tabular}{l|ccc}
\toprule
Shots & IBM Jakarta & IBM Lima & IBM Manila \\
\midrule
\midrule
512 & \textbf{0.0287} & 0.0266 & 0.0440 \\
\midrule
1024 & 0.0352 & 0.0246 & 0.0403 \\
\midrule
2048 & 0.0305 & \textbf{0.0217} & 0.0410 \\
\midrule
4096 & 0.0294 & 0.0250 & \textbf{0.0399} \\
\bottomrule

\end{tabular}
\label{tab:rmse_vs_shot}
\vspace{-15pt}

\end{table}

Table~\ref{tab:rmse_vs_layer} shows the relationship between the number of transformer layers with the prediction performance. We find that different model sizes do not greatly impact accuracy. On the noisy simulator and IBM Hanoi datasets, the one-layer model slightly outperforms the others, while on the IBM Geneva dataset, the three-layer model is the best. Therefore, in most of our experiments, we use a two-layer model as a trade-off.

Furthermore, we show the performance differences under different numbers of shots in noisy simulators as in Table~\ref{tab:rmse_vs_shot}. As the shots increase, the precision of the ground truth PST in the training set will be improved and will converge to the true PST when the shots are infinity. However, counter-intuitively, we find that increasing shot number does not guarantee better model accuracy.

Finally, besides theoretical proof of lower computation complexity of our model, we also perform empirical runtime comparisons as shown in Table~\ref{tab:runtime}. We run both the circuit simulator and the graph transformer on an Nvidia 3090 GPU with 24GB memory for 1000 sampled circuits from the random circuit dataset, and report average runtime. We select batch size 1 or 10 for the graph transformer predictor. The predictor achieves 200\x and 1.7K\x speedup over classical simulators to obtain the PST for batch size 1 and 10, respectively. That demonstrates the much higher efficiency of our graph transformer-based predictor.

\begin{table}[t]
\centering
\renewcommand*{\arraystretch}{1}
\setlength{\tabcolsep}{4pt}
\caption{Runtime of Simulation \vs Transformer Predictor}

\begin{tabular}{l|ccc}
\toprule
 & Simulation & Predictor (bsz=1) & Predictor (bsz=10) \\
 \midrule
 \midrule
Latency (s) & 5.57E-1 & 2.79E-3 & 3.28E-4 \\
\bottomrule

\end{tabular}
\label{tab:runtime}

\end{table}

\section{Conclusion}

Using machine learning to optimize quantum system problems is promising. This paper presents a case study of the ML for Quantum part of \tq library. 
We are inspired by that \textit{a quantum circuit is a graph} and propose to leverage a \textit{graph transformer} model to predict the circuit fidelity under the influence of quantum noise. 
First, we collect a large dataset of randomly generated circuits and algorithm circuits, and measure their fidelity on simulators and real machines.
A graph with feature vectors for each node is constructed according to the circuit. The graph transformer processes the circuit graph and calculates the anticipated fidelity value for the circuit.
Instead of the exponential cost of performing whole circuit simulations, we can effectively evaluate the fidelity under polynomial complexity. The datasets and models have been integrated into the \tq library, and we hope they can accelerate research in the ML and Quantum field.

\section*{Acknowledgment}

We thank National Science Foundation, MIT-IBM Watson AI Lab, and Qualcomm Innovation Fellowship for supporting this research. This work is funded in part by EPiQC, an NSF Expedition in Computing, under grants CCF-1730082/1730449; in part by STAQ under grant NSF Phy-1818914; in part by DOE grants DE-SC0020289 and DE-SC0020331; and in part by NSF OMA-2016136 and the Q-NEXT DOE NQI Center. We acknowledge the use of IBM Quantum services for this work.

\bibliographystyle{ACM-Reference-Format}
\bibliography{main}

\end{document}